\documentclass[a4paper,10pt,twoside]{cpc-hepnp}

\usepackage{multicol}
\usepackage{graphicx}
\usepackage{booktabs}
\usepackage{amssymb,bm,mathrsfs,bbm,amscd}
\usepackage[tbtags]{amsmath}
\usepackage{lastpage}
\usepackage{CJK}
\usepackage{color}
\usepackage{epstopdf}
\usepackage{epsfig}
\usepackage{ulem}
\usepackage{pdfsync}

\begin{document}
\begin{CJK*}{GB}{gbsn}

\title{Criticality of QCD in a holographic QCD model with critical end point}

\author{%
      Xun Chen(³ÂÑ«)$^{1,2;1)}$\email{chenxun@ihep.ac.cn}%
\quad Danning Li(ÀÄý)$^{3;2)}$\email{lidanning@jnu.edu.cn}%
\quad Mei Huang(»Æ÷)$^{2,4,5;3)}$\email{huangm@ihep.ac.cn}
}
\maketitle

\address{%
$^1$ Central China Normal University, Wuhan 430079, P.R. China\\
$^2$ Institute of High Energy Physics, Chinese Academy of Sciences, Beijing 100049, P.R. China\\
$^3$ Department of Physics and Siyuan Laboratory, Jinan University, Guangzhou 510632, P.R. China\\
$^4$ University of Chinese Academy of Sciences, Beijing 100049, P.R. China\\
$^5$ Theoretical Physics Center for Science Facilities, Chinese Academy of Sciences, Beijing 100049, P.R. China\\
}

\begin{abstract}
Thermodynamics of strongly interacting matter near critical end point are investigated in a holographic QCD model, which could describe the QCD phase diagram in $T-\mu$ plane qualitatively. Critical exponents along different axis ($\alpha,\beta,\gamma,\delta$) are extracted numerically. It is given that $\alpha\approx 0, \beta\approx 0.54, \gamma \approx 1.04, \delta \approx 2.97$, the same as 3D Ising mean-field approximation and previous holographic QCD model calculations. We also discuss the possibilities to go beyond mean field approximation by including the full back-reaction of chiral dynamics in the holographic framework.
\end{abstract}


\begin{multicols}{2}

\section{Introduction}\label{sec-int}
The critical phenomena, taking place in second or higher order phase transitions, displays several novel properties(e.g. universal power law divergence of some quantities, scaling relations of different quantities, and so on), which are strongly related to the correlations of fluctuations and cannot be described by mean field calculations at spatial dimension lower than $4$\cite{Pelissetto:2000ek}. It is firstly observed in carbon dioxide system experimentally\cite{critical1st}, and well studied in other systems(see Ref.\cite{Fisher:1974uq,Hohenberg:1977ym} for more details). In nuclear physics area, the progresses of Relativistic Heavy-Ion Collisions\cite{Aggarwal:2010cw,Odyniec:2013aaa,Luo:2017faz} have motivated a lot of interests in studying criticality of the critical end point(CEP) in Quantum Chromodynamics(QCD) phase transition(please refer to Refs.\cite{Stephanov:2004wx,Stephanov:2007fk} for details). It is widely accepted that with  physical quark masses QCD phase transition should be a continuous crossover at small baryon chemical potential, while it becomes a first order one at sufficient large baryon number density. In between the two kinds, a second order phase transition would occur at the CEP. Besides the criticality at CEP in $T-\mu$ plane with physical quark masses, the criticality of chiral critical lines in the whole quark masses plane $m_{u/d}-m_s$ is also attracted a lot of attentions\cite{chiral-th,columbia-plot-origin,Ding:2015ona,columbia-plot-figure,Endrodi:2007gc,Fisher:1974uq,Hohenberg:1977ym} theoretically.

Generally, the main properties of a critical point are described by a set of critical exponents, related to the leading scaling behavior of certain quantities along certain axes. For example, in common convention, the standard thermodynamical critical exponent $(\alpha,\beta,\gamma,\delta)$ are defined as follows. The exponent $\alpha$ is related to the leading power scaling of specific heat $C_\rho$ by
\begin{eqnarray}
C_\rho \sim|T-T_c|^{-\alpha},\label{cealpha}
\end{eqnarray}
as the critical point is approached along the first order axis with a certain density $\rho$. The exponent $\beta$ is defined as \begin{eqnarray}
\Delta\rho \sim(T_c-T)^{-\beta},\label{cegamma}
\end{eqnarray}
when the temperature $T$ is getting fairly close to the critical temperature $T_c$ along the first-order line. Here $\Delta\rho$ is the the discontinuity of $\rho$ across the first-order line. The exponent $\gamma$ is defined as
\begin{eqnarray}
\chi_2 \sim(T-T_c)^{-\gamma},
\end{eqnarray}
along the first-order axis also, with $\chi_2\equiv (\frac{\partial\rho}{\partial\mu})_{_T}$ the baryon number susceptibility. The exponent $\delta$ is defined as
\begin{gather}
\rho-\rho_c \sim(\mu-\mu_c)^{\frac{1}{\delta}},
\end{gather}
when $\mu$ approaching $\mu_c$ with temperature fixed to the critical temperature $T_c$. Besides the definition from thermodynamical quantities, the critical exponents could also be defined through the order parameters like chiral condensate\cite{Ding:2015ona,Laermann:2003cv} and pion condensate\cite{Wang:2015bky}. Besides the above critical exponents related to the scaling behavior of static quantities, there are also critical exponents describing the dynamical evolution when a system approaching the critical point, which we will not discussed here. From renormalization group theory\cite{Wilson:1993dy}, the critical exponents are related to each other by the following equations:
\begin{gather}
\alpha+2\beta+\gamma=2,\alpha+\beta(1+\delta)=2.
\end{gather}

The mean field calculation show that $\alpha=0,\beta=\frac{1}{2},\gamma=1,\delta=3$ in 3D. To study the critical behavior of QCD, many efforts have been paid to go beyond mean field approximation from different methods like lattice simulations\cite{AliKhan:2000wou,Ejiri:2009ac,Karsch:2010ya,Kaczmarek:2011zz,Burger:2011zc}, Dyson-Schwinger equations(DSEs)\cite{Fischer:2011pk,Fischer:2012vc}, functional renormalization group(FRG)\cite{Grahl:2014fna,Wang:2015bky}, $\epsilon$ expansion \cite{chiral-th,Yee:2017sir}. In general, the lattice simulations are the most reliable method to study QCD. It is extracted from lattice simulations that $\alpha\approx-0.213, \beta\approx 0.385, \gamma\approx 1.453, \delta\approx 4.824$ for $O(4)$ universality class\cite{Kanaya:1994qe,Engels:2001bq,Engels:2003nq} and $\alpha\approx 0.110, \beta\approx 0.327, \gamma\approx 1.237, \delta\approx 4.789$ for $Z(2)$ universality class\cite{Campostrini:2002cf}. But when quark masses become small, lattice simulations become fairly expensive. Moreover, the sign problem makes it hard to get reliable result at large baryon number density. Thus, it is quite necessary to investigate QCD phase transition from other methods, trying to probe the real properties of QCD. The development of anti-de Sitter/conformal field theory correspondence (AdS/CFT)\cite{Maldacena:1997re,Gubser:1998bc,Witten:1998qj} provides a new way to deal with strong interacting system like QCD. Many efforts have been made in building a realistic holographic QCD model to describe hadron physics\cite{Erlich:2005qh,Karch:2006pv,TB:05,DaRold2005,D3-D7,D4-D6,SS-1,SS-2,Csaki:2006ji,Dp-Dq,Gherghetta-Kapusta-Kelley,Gherghetta-Kapusta-Kelley-2,YLWu,YLWu-1,Li:2012ay,Li:2013oda,Bartz:2014oba,Colangelo:2008us,Bellantuono:2015fia,Capossoli:2013kb,Capossoli:2015ywa,Capossoli:2016kcr,Capossoli:2016ydo,Chen:2015zhh,Vega:2016gip} , hot/dense QCD matter\cite{Shuryak:2004cy,Tannenbaum:2006ch,Policastro:2001yc,Cai:2009zv,Cai:2008ph,Sin:2004yx,Shuryak:2005ia,Nastase:2005rp,Janik:2005zt,Nakamura:2006ih,Sin:2006pv,Herzog:2006gh,Gubser-drag,Wu:2014gla,Li:2014dsa,Li:2014hja,Li:2011hp,Cai:2012xh}, and so on. For near critical point physics in bottom-up holographic QCD, several groups have developed different models with CEP in the Einstein-Maxwell-Dilaton system\cite{Cai:2012xh,DeWolfe:2010he,DeWolfe:2011ts,Knaute:2017opk,Critelli:2017euk}. The static critical scaling near the CEP is investigated in \cite{DeWolfe:2010he}, and it is shown to be $\alpha=0,\beta\approx0.482,\gamma\approx0.942,\delta\approx3.035$, very close to the mean field results. Meanwhile, the dynamical critical exponents, which are related to dynamical evolution of the system towards the critical point, are analyzed in \cite{DeWolfe:2011ts,Knaute:2017opk,Critelli:2017euk}. Besides, the critical exponents at chiral critical lines are given in\cite{Li:2016smq}. Most of the studies gave a group of thermodynamical critical exponents very close to the mean field results. Probably, it might be connected to the suppression of quantum corrections by large $N_c$ as pointed out in \cite{DeWolfe:2010he}. In phenomenological sense, it is valuable to see whether one can describe the critical behavior beyond mean field level. Since if one tries to compare the holographic data to lattice simulations or experimental data, the value of $N_c$ is not infinite in some sence. In this work, we will follow the studies in \cite{Yang:2014bqa,Li:2017ple} and try to extract the critical exponents in a holographic QCD model with a critical end point.

The paper is organized as follows. In Sec.1, we give a brief introduction about the critical phenomena and holographic method. In Sec.2, we describe the Einstein-Dilaton-Maxwell system, which we will take as our starting point. In Sec.3, we will show the results of our holographic QCD model. Finally, in Sec.4, a short summary will  be given.

\section{Einstein-Maxwell-Dilaton system and critical end point of QCD phase diagram}\label{sec-action}

As mentioned above, the Einstein-Maxwell-Dilaton (EMD) system provides a good starting point to consider both finite temperature and finite chemical potential. By including a $U(1)$ gauge field, one can introduce the chemical potential to the system. In the framework of EMD system, the authors of \cite{Yang:2014bqa} proposed a holographic QCD model with a critical end point at $T_c=0.121 \rm{GeV}, \mu^B_c=0.693\rm{GeV}$. The model is shown to produce correct vector meson spectra as well as thermodynamical data. In \cite{Li:2017ple}, it is shown that the baryon number susceptibilities extracted in this model is comparable with experimental data.   Furthermore, the study shows the close relationship between the location of the critical end point and the location of the peak in baryon number susceptibilities. In this work, we will try to examine the critical behavior near the critical end point and check whether it is possible to go beyond mean filed level.

Firstly, for the compactness of this paper, we will briefly introduce the EMD system. Following\cite{Yang:2014bqa}, the action is taken as

\begin{gather}
S=S_b+S_m,\\
S_b=\frac{1}{16\pi G_5}\int d^5x\sqrt{-g}[R - \frac{f(\phi)}{4}F^2-\frac{1}{2}\partial_\mu\phi\partial^\mu\phi-V(\phi)]\label{sb},\\
S_m=\frac{1}{16\pi G_5}\int d^5x\sqrt{-g}[\frac{f(\phi)}{4}(F^2_V+F^2_{\tilde{V}})].
\end{gather}

Here $S$ is the full action including the background part $S_b$ and the matter part $S_m$, $g$ is the determinant of metric $g_{\mu\nu}$, $G_5$ is the 5D Newton constant, $\phi$ is the dilaton field, $F, F_V, F_{\tilde{V}}$ are the strength tensor of gauge field dual to the baryon number current, isospin vector current, and axial-vector current respectively. Generally, if $F,F_V,F_{\tilde{V}}=0$, it is reduced to the zero chemical potential case. If $F\neq0$, one could introduce baryon number chemical potential in the system, while for $F_V\neq 0$ one might introduce the corresponding chemical potential related to isospin number. In the study here, we would focus on cases with finite baryon number density and set $S_m=0$. The EMD system could describe glue-dynamics well. In \cite{Cai:2012xh}, it is shown to give well description of phase diagram in heavy quark limit. Also, in Einstein-Dilaton system, it was shown that the glue ball spectral and the pure gluon thermodynamics could be well described\cite{Li:2013oda,Li:2011hp}. To take chiral dynamics into account, a direct way is to add the $S_m$ part. Besides, like \cite{Yang:2014bqa}, one can add chiral dynamics by adjusting the dilaton potential $V(\phi)$ carefully. In some sense, $S_m$ part is replaced by the potential terms. In this work, we will follow \cite{Yang:2014bqa} and take the latter way.

To consider gravity dual to QCD, we will take the following metric ansatz

\begin{gather}
ds^2=\frac{e^{2A(z)}}{z^2}[-h(z)dt^2+\frac{1}{h(z)}dz^2+d\vec{x}^2],
\end{gather}

and consider the black hole solution. Inserting the metric ansatz into Eq.(\ref{sb}), one can derive the Einstein equation and simplify it as\cite{Yang:2014bqa}

\begin{gather}
\phi^{\prime\prime}+\left(  \frac{g^{\prime}}{g}+3A^{\prime}-\dfrac{3}%
{z}\right)  \phi^{\prime}+\left(  \frac{z^{2}e^{-2A}A_{t}^{\prime2}f_{\phi}%
}{2g}-\frac{e^{2A}V_{\phi}}{z^{2}g}\right)     =0,\label{eom-phi}\\
A_{t}^{\prime\prime}+\left(  \frac{f^{\prime}}{f}+A^{\prime}-\dfrac{1}%
{z}\right)  A_{t}^{\prime}    =0,\label{eom-At}\\A^{\prime\prime}-A^{\prime2}+\dfrac{2}{z}A^{\prime}+\dfrac{\phi^{\prime2}}{6}
 =0,\label{eom-A}\\
g^{\prime\prime}+\left(  3A^{\prime}-\dfrac{3}{z}\right)  g^{\prime}%
-e^{-2A}z^{2}fA_{t}^{\prime2}    =0,\label{eom-g}\\
A^{\prime\prime}+3A^{\prime2}+\left(  \dfrac{3g^{\prime}}{2g}-\dfrac{6}%
{z}\right)  A^{\prime}-\dfrac{1}{z}\left(  \dfrac{3g^{\prime}}{2g}-\dfrac
{4}{z}\right)  +\dfrac{g^{\prime\prime}}{6g}+\frac{e^{2A}V}{3z^{2}g}    =0.\label{eom-V}
\end{gather}

In the above equations, there are undefined functions like $A(z), A_t(z), f(z),\phi(z), V(\phi)$, $f(\phi)$, which should be input as the starting point of the model. Generally, there are different kinds of ways to deal with this issue as discussed in \cite{Li:2013oda}. One can input $V(\phi), f(\phi)$ as in Refs.\cite{DeWolfe:2010he} and solve $A(z), \phi(z),A_t(z)$ numerically from the equations of motion. In another way, one can input $\phi(z),f(\phi)$ or $A(z), f(\phi)$ and solve the rest, which is usually called 'potential reconstruction approach' and used in several works\cite{Cai:2012xh,Li:2011hp,He:2013qq,Yang:2014bqa,Li:2017tdz,Chen:2017cyc,Kajantie:2011nx}. As analyzed in \cite{Fang:2015ytf}, the potential reconstruction approach could be considered as a good approximation of the fixing potential method. The qualitative picture of the thermodynamical quantities and phase transition structure are the same as in fixing potential method. Since it is convenient and the main features are kept, we will follow the study in \cite{Yang:2014bqa} and use the potential reconstruction method.

\begin{figure*}
    \centering
    \includegraphics[width=10cm]{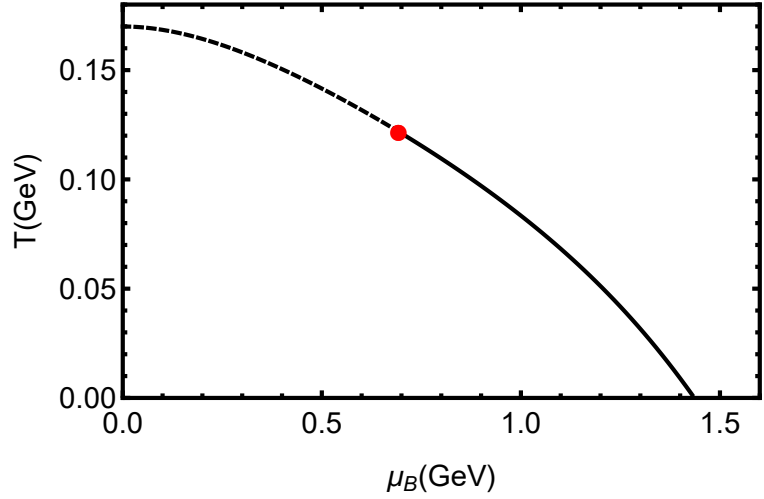}
    \caption{\label{cepline}The phase diagram in $T-\mu_B$ plane from the holographic QCD model defined in Eqs.(\ref{eqa},\ref{eqf}) with $b=-6.25\times 10^{-4} {\rm GeV}^4, c=0.227{\rm GeV}^2$. At small chemical potential, the phase transition is a crossover(the black dashed line), while it turns to be first order one(the black solid line) at large chemical potential. The critical end point(the red dot) locates at $(T^c,\mu_B^c)=(0.121{\rm GeV},0.693{\rm GeV})$.}
\end{figure*}

As in \cite{Yang:2014bqa}, we will take $A(z)$ and $f(z)$ as the form of
\begin{eqnarray}
A(z)=-\frac{c}{3} z^2-b z^4,\label{eqa}\\
f(\phi(z))=e^{c z^2-A(z)}.\label{eqf}
\end{eqnarray}
Here, $b,c$ are model parameters, which are fixed from the meson spectra and sound speed data to be $b=-6.25\times 10^{-4} {\rm GeV}^4, c=0.227{\rm GeV}^2$. Requiring the boundary conditions at the horizon $z=z_H$ and boundary $z=0$
\begin{eqnarray}
A_t(z_H)=g(z_H)=0,\\
A(0)=-\sqrt{\frac{1}{6}}\phi(0), \hspace{0.3cm} g(0)=1,\\
A_t(0)=\mu+\rho z^2+\cdots,
\end{eqnarray}
one can solve the rest unknown functions, which are described in detail in \cite{Yang:2014bqa} and we will not repeat here. To study the thermodynamical properties, one can extract the baryon number density $\rho$, entropy density $s$, temperature $T$ from the background solution as
\begin{eqnarray}
\rho=\frac{c \mu }{1-e^{c z_H^2}},\\
s=2\pi \frac{e^{3A(z_H)}}{z_H^3},
\end{eqnarray}
\begin{gather}
T=\frac{z_H^3 e^{-3 A(z_H)}}{4 \pi  \int_0^{z_H} y^3 e^{-3 A(y)} \, dy}[1-\frac{2 c \mu ^2}{(1-e^{c z_H^2})^2}\times
  \nonumber \\
 (e^{c z_H^2} \int_0^{z_H} y^3 e^{-3 A(y)} \, dy-\int_0^{z_H} y^3 e^{c y^2-3 A(y)} \, dy)].
\end{gather}
Starting from the above equations, one can reach the free energy by thermodynamical relations $F=-\int [sdT+\rho d\mu]$. It is found that at small chemical potential region, there is only one solution for each temperature. The low energy phase and high energy phase are connected smoothly, which gives a crossover transition. Meanwhile, when the chemical potential is sufficiently large, there would be three branches of solutions in a short temperature region, showing a typical first order transition. Minimizing the free energy, one can determine the transition temperature of the first order phase transition. The phase transition line was determined as in Fig.\ref{cepline}, where we find a crossover line(the black dashed line) and a first order line(the black solid line). The boundary of the crossover line and the first order line is the critical end point,  which locates at $(T^c,\mu_B^c)=(0.121{\rm GeV},0.693{\rm GeV})$. In next section, we will focus on studying the near critical point behavior of this CEP.

\section{Holographic results of critical exponents at CEP}\label{sec-res}
Comparing to the traditional method, holographic methods describe the system in a higher dimension. The fifth dimension in holographic QCD could be mapped to energy scale and the evolution of fields in fifth dimension could be mapped to running of couplings in 4D theory. In this sense, one should expect the corrections from holographic methods to critical exponents given by mean field theory. However, the studies in \cite{DeWolfe:2010he} showed that the corrections are suppressed in holographic method. In order to check this point, we would study the near critical point behavior in an independent holographic model. More concretely, in this section, we would focus on extracting the critical exponents $(\alpha, \beta, \gamma, \delta)$ of holographic model described in last section.

\begin{figure*}
    \centering
    \includegraphics[width=10cm]{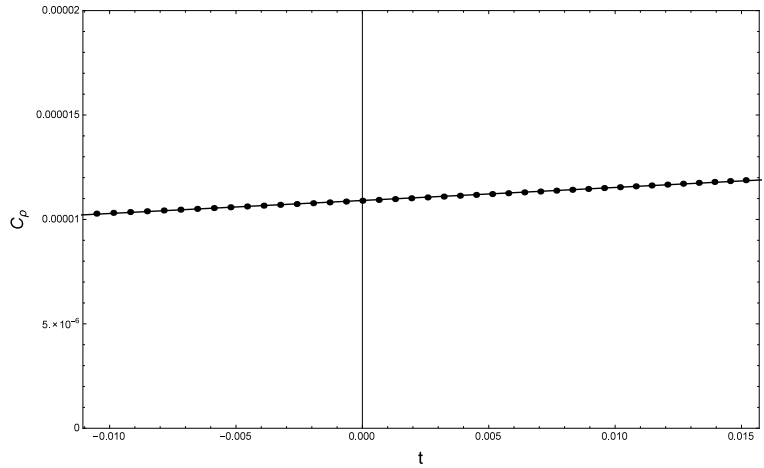}
    \caption{\label{alpha}The specific heat $C_\rho$ around the critical point along a line of constant $\rho$, along the
first-order axis. From this figure, we can see $\alpha$ = 0.}
\end{figure*}

Firstly, we would analyze the specific heat, which is defined as $C_\rho\equiv T(\frac{\partial s}{\partial T})_{_\rho}$. We fix the baryon number density and take the derivative of entropy density with respect to $T$. We find that $C_\rho$ is a regular function of temperature at $T=T_c$ as shown in Fig.\ref{alpha}. This indicates that the exponent $\alpha$ defined in Eq.(\ref{cealpha}) would be
\begin{eqnarray}
\alpha = 0.
\end{eqnarray}

\begin{figure*}
\begin{center}
\epsfxsize=6.5 cm \epsfysize=6.5 cm \epsfbox{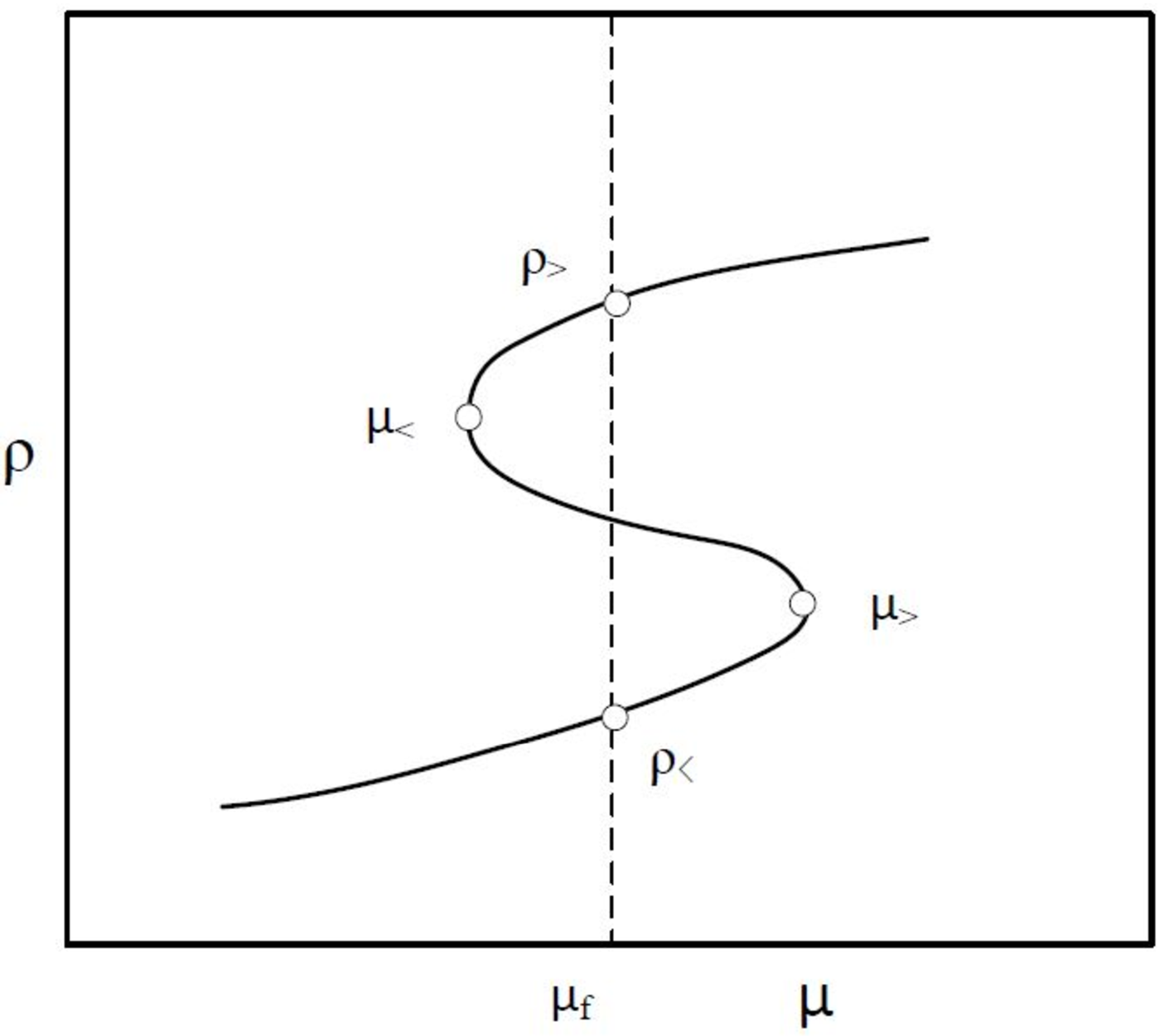}
\hspace*{0.1cm} \epsfxsize=6.5 cm \epsfysize=6.5 cm
\epsfbox{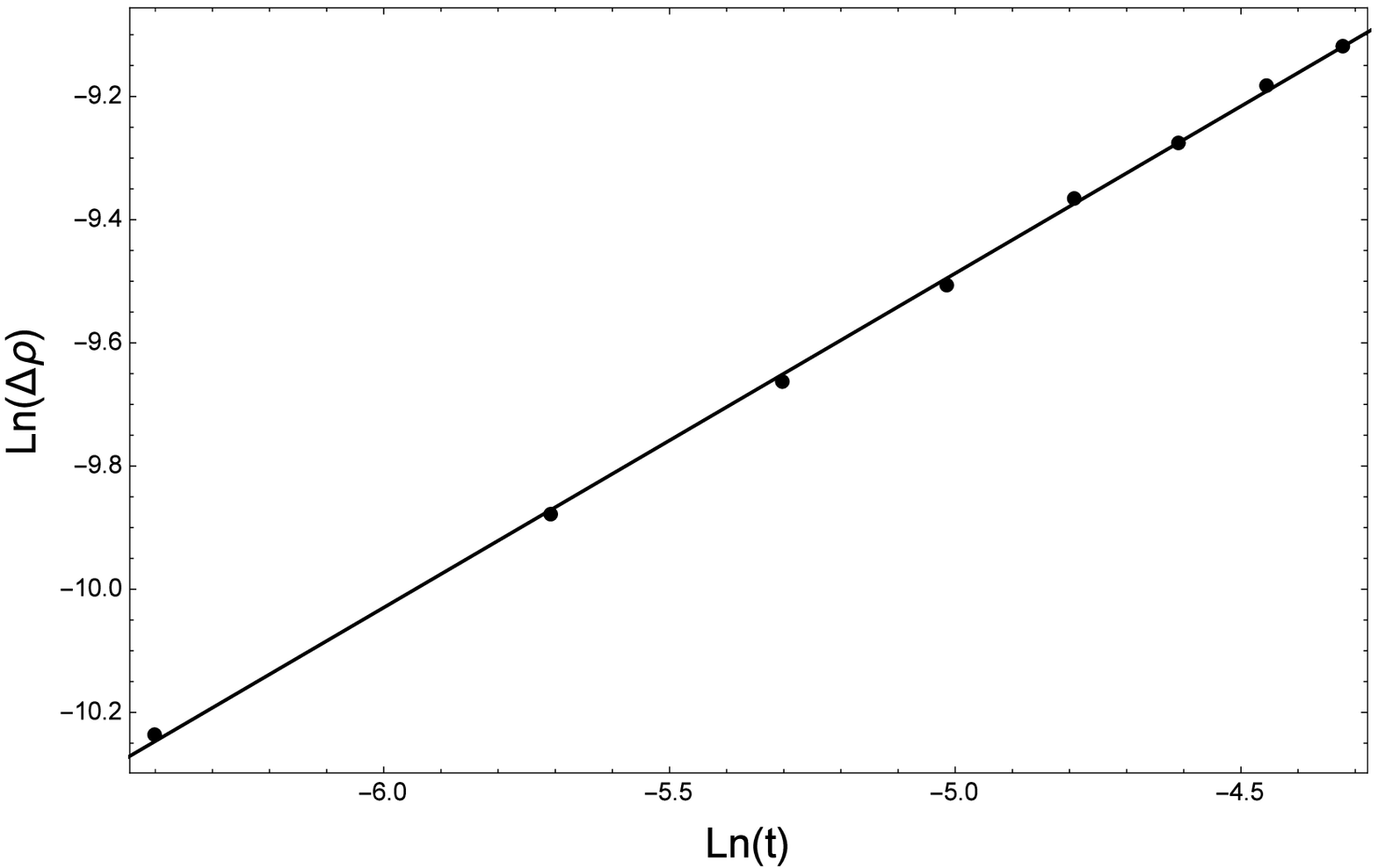} \vskip -0.05cm \hskip 0.15 cm
\textbf{( a ) } \hskip 6.5 cm \textbf{( b )} \\
\end{center}
\caption{The discontinuity in the baryon number density on a
log-log plot.The slope gives us a value $\beta$ = 0.54, here $t\equiv(T-T_c)/T_c$
 }
 \label{beta}
\end{figure*}

Then, we would examine the critical behavior of discontinuity of baryon number density $\Delta \rho$. It is easy to understand that near the crossover line in Fig.\ref{cepline} the difference of $\rho$ of the two sides should be zero. So we will focus on the first order line. Near the first order line, at a certain temperature, the density $\rho$ are triple-valued function of the baryon chemical potential, as shown in Fig.\ref{beta}(a). From Fig.\ref{beta}(a), we could see that outside the range $\mu_< <\mu<\mu_>$ the solution of equation of motion is single-valued, while inside this region it is triple-valued. Since the critical exponent $\beta$ is defined along the first order line, we have to determine the transition chemical potential. In principle, it should be done by minimizing the free energy, or equivalently from Maxwell's equal-area construction.  Here, we use Maxwell's equal-area construction to determinate transition point. Considering that we only care about the critical exponent of $\Delta\rho$, we can simply consider $\Delta\rho \equiv\rho_>-\rho_<$ on the first order line. Under this convention, we show the log-log plot of $(\log(t),\log(\Delta\rho))$ in Fig.\ref{beta}(b), with $t \equiv(T-T_c)/T_c$. From this plot, we could see that the data points of $(\log(t),\log(\Delta\rho))$ lie almost in a straight line, showing the power low divergence of the critical behavior of $\Delta\rho$ versus $T-T_c$. From the linear fitting, we find that the slope of the straight line is about $0.54$, which means
\begin{eqnarray}
\beta\approx 0.54,
\end{eqnarray}
very close to the mean field value $\beta=\frac{1}{2}$ in 3D.

\begin{figure*}
    \centering
    \includegraphics[width=10cm]{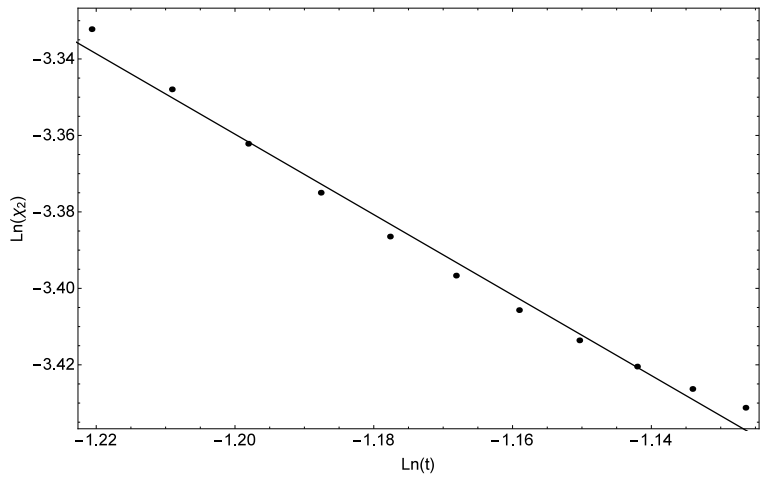}
    \caption{\label{gamma}{The baryon number susceptibility $\chi_2$ vs t $\equiv$ ($T-T_c$)/$T_c$ as the critical point is
approached on a log-log plot. The slope of a best fit line through the data can give us a value
 $\gamma= 1.05$.}}
\end{figure*}

As defined in Eq.(\ref{cegamma}), the critical exponent comes from the baryon number susceptibility $\chi_2\equiv (\frac{\partial\rho}{\partial\mu})_{_T}$. Taking the derivative of $\rho$ with respective to $\mu$, we can extract $\chi_2$ from the results of $\rho$. The log-log plot of $\chi_2$ versus $T$ is shown in Fig.\ref{gamma}. From the plot, we could see that $\chi_2$ diverges at the critical end point $t=0$. All the data points of $(\log(t),\log(\chi_2))$ lie almost on a straight line. From the linear fitting, we get the slope of the straight line is $0.95$, which implies the critical exponent $\gamma=1.05$.

\begin{figure*}
    \centering
    \includegraphics[width=10cm]{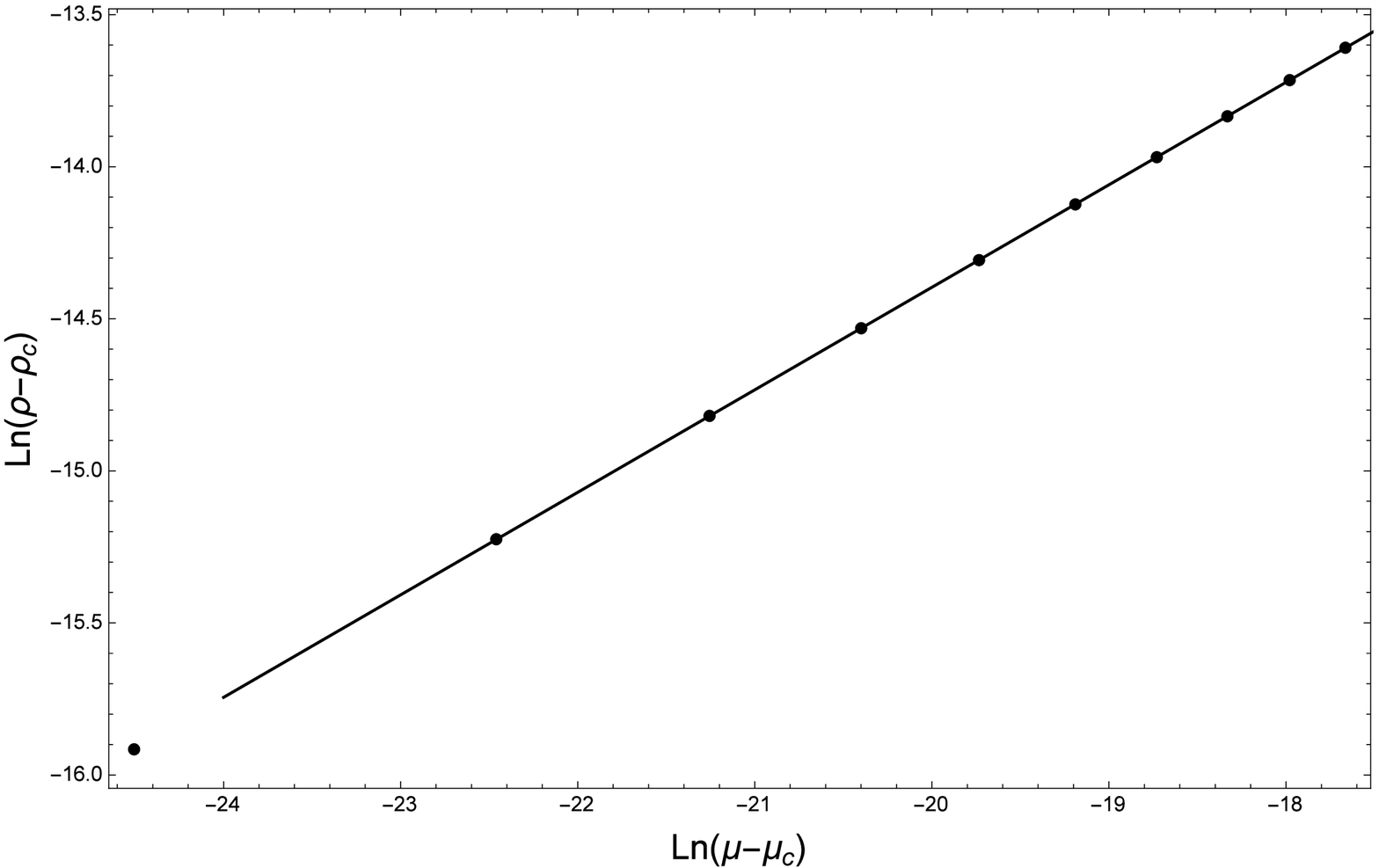}
    \caption{\label{delta}$\rho$ approaches $\rho_c$ as $\mu$ approaches $\mu_c$ on the critical isotherm on
a log-log plot. The slope shows $\delta$ = 2.97.}
\end{figure*}

Taking $T=T_c=0.121\rm{GeV},\mu=\mu_c=0.693\rm{GeV}$, one can get $\rho_c=0.00004858 $. Keeping $T=T_c$ and changing $\mu$, one finds $\rho$ would changes correspondingly.  The log-log plot of $\rho-\rho_c$ versus $\mu-\mu_c$ is given in Fig.\ref{delta}. There we see that all the data points of $(\log(\mu-\mu_c),\log(\rho-\rho_c))$ lie almost in a straight line, showing the leading power low of  the critical behavior of $\rho-\rho_c$ versus $\mu-\mu_c$. From the linear fitting, we find that the slope of the straight line is about $0.337$, which means
\begin{eqnarray}
\delta\approx 2.97,
\end{eqnarray}
very close to the mean field value $\delta=3$ in 3D.

\end{multicols}

\begin{multicols}{2}
\section{Conclusion}\label{sec-sum}

We study critical behavior of hot and dense QCD matter in the holographic QCD model from \cite{Yang:2014bqa}, which could describe both the meson spectra quantitatively and phase diagram qualitatively. We extracted the specific heat $C_\rho$, discontinuity of baryon number density $\Delta\rho$, baryon number sucesptibility $\chi_2$ and baryon number density $\rho$ near the critical point. We find that the specific heat is a regular function of $T$ near the CEP, which indicates $\alpha=0$. Different from the specific heat, the other three quantities diverge at the critical end point.  We show the log-log plot of these quantities and find that the numerical data lies almost in straight lines for the three quantities, which indicates the leading power law of the critical scaling. From linear fitting, we reach the results $\alpha=0,\beta\approx 0.54, \gamma=1.05, \delta=2.97 $, almost the same as the 3D Ising mean field result, which shares the same critical exponents with a large variety of systems in mean field level.

The result is coincident with previous study in different holographic models \cite{DeWolfe:2011ts}. The current result is within the mean field approximation. However, in some sense the holographic method could characterize the running coupling of 4D theory. One of the possible reason might be due to the large $N_c$ suppression of the quantum corrections as mentioned in \cite{DeWolfe:2010he}. If we consider this question from a more general Gauge/Gravity duality, going outside the limitation of the large $N_c$ from a phenomenological way, there might be another possibility. The EDM system might describe the pure glue system only. Though if one adjusts the potential $V(\phi)$ or other quantities carefully to incooperate the effects of chiral dynamics, only the interaction far away from the critical end point are well described. Near the critical end point, due to the divergence of correlation length, the contribution from chiral dynamics might also be important\footnote{One can recall that the pure glue phase structure is totally different from that with physical quarks.}. Therefore, we might expect the correct couplings between glue-dynamics and chiral dynamics could improve this problem. Actually, in our recent study in \cite{criticalchiral}, we found that the running of dilaton field with temperature would affect the chiral critical exponents in soft-wall model. If the dilaton field does not depend on temperature, the chiral critical exponents would be mean field result, while when dilaton field exhibit critical scaling near critical temperature, the chiral critical exponents would go beyond mean field level (For more details, please refer to \cite{criticalchiral}.) Therefore, we expect that if we correctly couple the two dynamics in a full system, the correct critical behavior might be reached.

\section{Acknowledgments}

\acknowledgments{This work is supported by the NSFC under Grant No. 11725523, 11735007, 11805084 and 11261130311(CRC 110 by DFG and NSFC)}

\end{multicols}

\vspace{10mm}

\begin{multicols}{2}

\end{multicols}

\vspace{-1mm}
\centerline{\rule{80mm}{0.1pt}}
\vspace{2mm}

\begin{multicols}{2}

\end{multicols}

\clearpage

\end{CJK*}

\begin{thebibliography}{90}

\vspace{3mm}

\bibitem{Pelissetto:2000ek}
  A.~Pelissetto and E.~Vicari,
  ``Critical phenomena and renormalization group theory,''
  Phys.\ Rept.\  {\bf 368} (2002) 549

\bibitem{critical1st}
  B.~Berche, M.~Henkel, R.~Kenna,
  ``Critical phenomena: 150 years since Cagniard de la Tour,''
   Journal of Physical Studies 13 (3)(2009) , pp. 3001-1-3001-4.


\bibitem{Fisher:1974uq}
  M.~E.~Fisher,
  ``The renormalization group in the theory of critical behavior,''
  Rev.\ Mod.\ Phys.\  {\bf 46} (1974) 597
   Erratum: [Rev.\ Mod.\ Phys.\  {\bf 47} (1975) 543].

\bibitem{Hohenberg:1977ym}
  P.~C.~Hohenberg and B.~I.~Halperin,
  ``Theory of Dynamic Critical Phenomena,''
  Rev.\ Mod.\ Phys.\  {\bf 49} (1977) 435.


\bibitem{Aggarwal:2010cw}
  M.~M.~Aggarwal {\it et al.} [STAR Collaboration],
  ``An Experimental Exploration of the QCD Phase Diagram: The Search for the Critical Point and the Onset of De-confinement,''
  arXiv:1007.2613 [nucl-ex].

\bibitem{Odyniec:2013aaa}
  G.~Odyniec,
  ``RHIC Beam Energy Scan Program: Phase I and II,''
  PoS CPOD {\bf 2013} (2013) 043.

\bibitem{Luo:2017faz}
  X.~Luo and N.~Xu,
  ``Search for the QCD Critical Point with Fluctuations of Conserved Quantities in Relativistic Heavy-Ion Collisions at RHIC : An Overview,''
  Nucl.\ Sci.\ Tech.\  {\bf 28} (2017) no.8,  112
  [arXiv:1701.02105 [nucl-ex]].

\bibitem{Stephanov:2004wx}
  M.~A.~Stephanov,
  ``QCD phase diagram and the critical point,''
  Prog.\ Theor.\ Phys.\ Suppl.\  {\bf 153}, 139 (2004)
  [Int.\ J.\ Mod.\ Phys.\ A {\bf 20}, 4387 (2005)].

  \bibitem{Stephanov:2007fk}
  M.~A.~Stephanov,
  ``QCD phase diagram: An Overview,''
  PoS LAT {\bf 2006}, 024 (2006).

\bibitem{chiral-th}
  R.~D.~Pisarski and F.~Wilczek,
  ``Remarks on the Chiral Phase Transition in Chromodynamics,''
  Phys.\ Rev.\ D {\bf 29} (1984) 338.




\bibitem{columbia-plot-origin}
  F.~R.~Brown, F.~P.~Butler, H.~Chen, N.~H.~Christ, Z.~h.~Dong, W.~Schaffer, L.~I.~Unger and A.~Vaccarino,
  ``On the existence of a phase transition for QCD with three light quarks,''
  Phys.\ Rev.\ Lett.\  {\bf 65} (1990) 2491.

\bibitem{Ding:2015ona}
  H.~T.~Ding, F.~Karsch and S.~Mukherjee,
  ``Thermodynamics of strong-interaction matter from Lattice QCD,''
  Int.\ J.\ Mod.\ Phys.\ E {\bf 24} (2015) no.10,  1530007
  [arXiv:1504.05274 [hep-lat]].

\bibitem{columbia-plot-figure}
  E.~Laermann and O.~Philipsen,
  ``The Status of lattice QCD at finite temperature,''
  Ann.\ Rev.\ Nucl.\ Part.\ Sci.\  {\bf 53} (2003) 163
  [hep-ph/0303042].

\bibitem{Endrodi:2007gc}
  G.~Endrodi, Z.~Fodor, S.~D.~Katz and K.~K.~Szabo,
  ``The Nature of the finite temperature QCD transition as a function of the quark masses,''
  PoS LATTICE {\bf 2007} (2007) 182
  [arXiv:0710.0998 [hep-lat]].


\bibitem{Laermann:2003cv}
  E.~Laermann and O.~Philipsen,
  ``The Status of lattice QCD at finite temperature,''
  Ann.\ Rev.\ Nucl.\ Part.\ Sci.\  {\bf 53} (2003) 163
  [hep-ph/0303042].

\bibitem{Wang:2015bky}
  Z.~Wang and P.~Zhuang,
  ``Critical Behavior and Dimension Crossover of Pion Superfluidity,''
  Phys.\ Rev.\ D {\bf 94} (2016) no.5,  056012
  [arXiv:1511.05279 [hep-ph]].

\bibitem{Wilson:1993dy}
  K.~G.~Wilson,
  ``The renormalization group and critical phenomena,''
  Rev.\ Mod.\ Phys.\  {\bf 55} (1983) 583.




\bibitem{AliKhan:2000wou}
  A.~Ali Khan {\it et al.} [CP-PACS Collaboration],
  ``Phase structure and critical temperature of two flavor QCD with renormalization group improved gauge action and clover improved Wilson quark action,''
  Phys.\ Rev.\ D {\bf 63} (2001) 034502
  [hep-lat/0008011].

\bibitem{Ejiri:2009ac}
  S.~Ejiri {\it et al.},
  ``On the magnetic equation of state in (2+1)-flavor QCD,''
  Phys.\ Rev.\ D {\bf 80} (2009) 094505
  [arXiv:0909.5122 [hep-lat]].

\bibitem{Karsch:2010ya}
  F.~Karsch,
  ``O(N) universality and the chiral phase transition in QCD,''
  Prog.\ Theor.\ Phys.\ Suppl.\  {\bf 186}, 479 (2010)
  [arXiv:1007.2393 [hep-lat]].

\bibitem{Kaczmarek:2011zz}
  O.~Kaczmarek {\it et al.},
  ``Phase boundary for the chiral transition in (2+1) -flavor QCD at small values of the chemical potential,''
  Phys.\ Rev.\ D {\bf 83} (2011) 014504
  [arXiv:1011.3130 [hep-lat]].

\bibitem{Burger:2011zc}
  F.~Burger {\it et al.} [tmfT Collaboration],
  ``Thermal QCD transition with two flavors of twisted mass fermions,''
  Phys.\ Rev.\ D {\bf 87} (2013) no.7,  074508
  [arXiv:1102.4530 [hep-lat]].

\bibitem{Fischer:2011pk}
  C.~S.~Fischer and J.~A.~Mueller,
  ``On critical scaling at the QCD $N_f=2$ chiral phase transition,''
  Phys.\ Rev.\ D {\bf 84} (2011) 054013
  [arXiv:1106.2700 [hep-ph]].

\bibitem{Fischer:2012vc}
  C.~S.~Fischer and J.~Luecker,
  ``Propagators and phase structure of Nf=2 and Nf=2+1 QCD,''
  Phys.\ Lett.\ B {\bf 718} (2013) 1036
  [arXiv:1206.5191 [hep-ph]].

\bibitem{Grahl:2014fna}
  M.~Grahl,
  ``$U(2)_A\times U(2)_V$-symmetric fixed point from the functional renormalization group,''
  Phys.\ Rev.\ D {\bf 90}, no. 11, 117904 (2014)
  [arXiv:1410.0985 [hep-th]].


\bibitem{Yee:2017sir}
  H.~U.~Yee,
  ``Dynamic universality class of model H with frustrated diffusion: $\epsilon$ expansion,''
  Phys.\ Rev.\ D {\bf 97} (2018) no.1,  016003
  [arXiv:1707.08560 [hep-ph]].

\bibitem{Kanaya:1994qe}
  K.~Kanaya and S.~Kaya,
  ``Critical exponents of a three dimensional O(4) spin model,''
  Phys.\ Rev.\ D {\bf 51} (1995) 2404
  [hep-lat/9409001].

\bibitem{Engels:2001bq}
  J.~Engels, S.~Holtmann, T.~Mendes and T.~Schulze,
  ``Finite size scaling functions for 3-d O(4) and O(2) spin models and QCD,''
  Phys.\ Lett.\ B {\bf 514} (2001) 299
  [hep-lat/0105028].

\bibitem{Engels:2003nq}
  J.~Engels, L.~Fromme and M.~Seniuch,
  ``Correlation lengths and scaling functions in the three-dimensional O(4) model,''
  Nucl.\ Phys.\ B {\bf 675} (2003) 533
  [hep-lat/0307032].

\bibitem{Campostrini:2002cf}
  M.~Campostrini, A.~Pelissetto, P.~Rossi and E.~Vicari,
  ``25th order high temperature expansion results for three-dimensional Ising like systems on the simple cubic lattice,''
  Phys.\ Rev.\ E {\bf 65} (2002) 066127
  [cond-mat/0201180].






\bibitem{Maldacena:1997re}
  J.~M.~Maldacena,
  ``The large N limit of superconformal field theories and supergravity,''
  Adv.\ Theor.\ Math.\ Phys.\  {\bf 2}, 231 (1998)
  [Int.\ J.\ Theor.\ Phys.\  {\bf 38}, 1113 (1999)]  [arXiv:hep-th/9711200].

\bibitem{Gubser:1998bc}
  S.~S.~Gubser, I.~R.~Klebanov and A.~M.~Polyakov,
  ``Gauge theory correlators from non-critical string theory,''
  Phys.\ Lett.\  B {\bf 428}, 105 (1998)
  [arXiv:hep-th/9802109].

\bibitem{Witten:1998qj}
  E.~Witten,
  ``Anti-de Sitter space and holography,''
  Adv.\ Theor.\ Math.\ Phys.\  {\bf 2}, 253 (1998)
  [arXiv:hep-th/9802150].


 \bibitem{Erlich:2005qh}
  J.~Erlich, E.~Katz, D.~T.~Son, M.~A.~Stephanov,
  ``QCD and a holographic model of hadrons,''
  Phys.\ Rev.\ Lett.\  {\bf 95}, 261602 (2005).
  [hep-ph/0501128].

\bibitem{Karch:2006pv}
  A.~Karch, E.~Katz, D.~T.~Son and M.~A.~Stephanov,
 ``Linear confinement and AdS/QCD,''
 Phys.\ Rev.\ D {\bf 74} (2006) 015005.


 \bibitem{TB:05} G.~F.~de Teramond and S.~J.~Brodsky,
``The hadronic spectrum of a holographic dual of QCD,''
Phys.\ Rev.\ Lett.\ \textbf{94}, 201601 (2005). 

\bibitem{DaRold2005} L.~Da Rold and A.~Pomarol,
``Chiral symmetry breaking from five dimensional spaces,''
Nucl.\ Phys.\ B \textbf{721}, 79 (2005). 

\bibitem{D3-D7} J.~Babington, J.~Erdmenger, N.~J.~Evans, Z.~Guralnik and
I.~Kirsch, ``Chiral symmetry breaking and pions in
non-supersymmetric gauge/gravity
Phys.\ Rev.\ D \textbf{69}, 066007 (2004) [arXiv:hep-th/0306018].

\bibitem{D4-D6}
 M.~Kruczenski, D.~Mateos, R.~C.~Myers and D.~J.~Winters,
  ``Towards a holographic dual of large N(c) QCD,''
  JHEP {\bf 0405} (2004) 041[hep-th/0311270].

\bibitem{SS-1} T.~Sakai and S.~Sugimoto,
``Low energy hadron physics in holographic QCD,'' Prog.\ Theor.\
Phys.\ \textbf{113}, 843 (2005)[arXiv:hep-th/0412141].

\bibitem{SS-2}
 T.~Sakai and S.~Sugimoto,
``More on a holographic dual of QCD,'' Prog.\ Theor.\ Phys.\
\textbf{114}, 1083 (2006).  [arXiv:hep-th/0507073].

\bibitem{Csaki:2006ji}
  C.~Csaki and M.~Reece,
  ``Toward a systematic holographic QCD: A braneless approach,''
  JHEP {\bf 0705}, 062 (2007)
  [arXiv:hep-ph/0608266].


\bibitem{Dp-Dq}
 S.~He, M.~Huang, Q.~S.~Yan and Y.~Yang,
  ``Confront Holographic QCD with Regge Trajectories,''
  Eur.Phys.J.C.(2010)66:187.
  arXiv:0710.0988 [hep-ph].

 \bibitem{Gherghetta-Kapusta-Kelley}
T.~Gherghetta, J.~I.~Kapusta and T.~M.~Kelley,
  ``Chiral symmetry breaking in the soft-wall AdS/QCD model,''
Phys.\ Rev.\ D {\bf 79} (2009) 076003; 

\bibitem{Gherghetta-Kapusta-Kelley-2}
T.~M.~Kelley, S.~P.~Bartz and J.~I.~Kapusta,
  ``Pseudoscalar Mass Spectrum in a Soft-Wall Model of AdS/QCD,''
  Phys.\ Rev.\ D {\bf 83} (2011) 016002;

\bibitem{YLWu}
  Y.~-Q.~Sui, Y.~-L.~Wu, Z.~-F.~Xie and Y.~-B.~Yang,
  ``Prediction for the Mass Spectra of Resonance Mesons in the Soft-Wall AdS/QCD with a Modified 5D Metric,''
  Phys.\ Rev.\ D {\bf 81} (2010) 014024;  

\bibitem{YLWu-1}
Y.~-Q.~Sui, Y.~-L.~Wu and Y.~-B.~Yang,
  ``Predictive AdS/QCD Model for Mass Spectra of Mesons with Three Flavors,''
  Phys.\ Rev.\ D {\bf 83} (2011) 065030.  



\bibitem{Li:2012ay}
  D.~Li, M.~Huang and Q.~S.~Yan,
  ``A dynamical soft-wall holographic QCD model for chiral symmetry breaking and linear confinement,''
  Eur.\ Phys.\ J.\ C {\bf 73} (2013) 2615
  [arXiv:1206.2824 [hep-th]].


\bibitem{Li:2013oda}
  D.~Li and M.~Huang,
  ``Dynamical holographic QCD model for glueball and light meson spectra,''
  JHEP {\bf 1311} (2013) 088
  [arXiv:1303.6929 [hep-ph]].

\bibitem{Bartz:2014oba}
  S.~P.~Bartz and J.~I.~Kapusta,
  ``Dynamical three-field AdS/QCD model,''
  Phys.\ Rev.\ D {\bf 90} (2014) no.7,  074034
  [arXiv:1406.3859 [hep-ph]].

\bibitem{Colangelo:2008us}
  P.~Colangelo, F.~De Fazio, F.~Giannuzzi, F.~Jugeau and S.~Nicotri,
  ``Light scalar mesons in the soft-wall model of AdS/QCD,''
   Phys.\ Rev.\ D {\bf 78}, 055009 (2008)  [arXiv:0807.1054 [hep-ph]].  

\bibitem{Bellantuono:2015fia}
  L.~Bellantuono, P.~Colangelo and F.~Giannuzzi,
  ``Holographic Oddballs,''
  JHEP {\bf 1510} (2015) 137
  [arXiv:1507.07768 [hep-ph]].

\bibitem{Capossoli:2013kb}
  E.~Folco Capossoli and H.~Boschi-Filho,
  ``Odd spin glueball masses and the Odderon Regge trajectories from the holographic hardwall model,''
  Phys.\ Rev.\ D {\bf 88} (2013) no.2,  026010
  [arXiv:1301.4457 [hep-th]].

\bibitem{Capossoli:2015ywa}
  E.~Folco Capossoli and H.~Boschi-Filho,
  Phys.\ Lett.\ B {\bf 753} (2016) 419
  [arXiv:1510.03372 [hep-ph]].


\bibitem{Capossoli:2016kcr}
  E.~Folco Capossoli, D.~Li and H.~Boschi-Filho,
  ``Pomeron and Odderon Regge Trajectories from a Dynamical Holographic Model,''
  Phys.\ Lett.\ B {\bf 760} (2016) 101
  [arXiv:1601.05114 [hep-ph]].

\bibitem{Capossoli:2016ydo}
  E.~Folco Capossoli, D.~Li and H.~Boschi-Filho,
  ``Dynamical corrections to the anomalous holographic soft-wall model: the pomeron and the odderon,''
  Eur.\ Phys.\ J.\ C {\bf 76} (2016) no.6,  320
  [arXiv:1604.01647 [hep-ph]].

\bibitem{Chen:2015zhh}
  Y.~Chen and M.~Huang,
  ``Two-Gluon and Trigluon Glueballs from Dynamical Holography QCD,''
  arXiv:1511.07018 [hep-ph].

\bibitem{Vega:2016gip}
  A.~Vega and P.~Cabrera,
  ``Family of dilatons and metrics for AdS/QCD models,''
  Phys.\ Rev.\ D {\bf 93} (2016) no.11,  114026
  [arXiv:1601.05999 [hep-ph]].



\bibitem{Shuryak:2004cy}
  E.~V.~Shuryak,
  ``What RHIC experiments and theory tell us about properties of  quark-gluon plasma?,''
  Nucl.\ Phys.\  A {\bf 750}, 64 (2005)
  [arXiv:hep-ph/0405066].

\bibitem{Tannenbaum:2006ch}
  M.~J.~Tannenbaum,
  ``Recent results in relativistic heavy ion collisions: From ' a new state of
  Rept.\ Prog.\ Phys.\  {\bf 69}, 2005 (2006)
  [arXiv:nucl-ex/0603003].

\bibitem{Policastro:2001yc}
  G.~Policastro, D.~T.~Son and A.~O.~Starinets,
  ``The shear viscosity of strongly coupled N = 4 supersymmetric Yang-Mills plasma,''
  Phys.\ Rev.\ Lett.\  {\bf 87}, 081601 (2001)
  [arXiv:hep-th/0104066].

\bibitem{Cai:2009zv}
  R.~-G.~Cai, Z.~-Y.~Nie, N.~Ohta and Y.~-W.~Sun,
  ``Shear Viscosity from Gauss-Bonnet Gravity with a Dilaton Coupling,''  Phys.\ Rev.\ D {\bf 79}, 066004 (2009)  [arXiv:0901.1421 [hep-th]].  
\bibitem{Cai:2008ph}
  R.~-G.~Cai, Z.~-Y.~Nie and Y.~-W.~Sun,
  ``Shear Viscosity from Effective Couplings of Gravitons,''  Phys.\ Rev.\ D {\bf 78}, 126007 (2008)  [arXiv:0811.1665 [hep-th]].  


\bibitem{Sin:2004yx}
  S.~J.~Sin and I.~Zahed,
  ``Holography of radiation and jet quenching,''
  Phys.\ Lett.\  B {\bf 608}, 265 (2005)
  [arXiv:hep-th/0407215];\\

\bibitem{Shuryak:2005ia}
  E.~Shuryak, S.~J.~Sin and I.~Zahed,
  ``A Gravity Dual of RHIC Collisions,''
  J.\ Korean Phys.\ Soc.\  {\bf 50}, 384 (2007)
  [arXiv:hep-th/0511199].

\bibitem{Nastase:2005rp}
  H.~Nastase,
  ``The RHIC fireball as a dual black hole,''
  arXiv:hep-th/0501068.


\bibitem{Janik:2005zt}
  R.~A.~Janik and R.~B.~Peschanski,
  ``Asymptotic perfect fluid dynamics as a consequence of AdS/CFT,''
  Phys.\ Rev.\  D {\bf 73}, 045013 (2006)
  [arXiv:hep-th/0512162];

\bibitem{Nakamura:2006ih}
  S.~Nakamura and S.~J.~Sin,
 ``A holographic dual of hydrodynamics,''
  JHEP {\bf 0609}, 020 (2006)
  [arXiv:hep-th/0607123];\\

\bibitem{Sin:2006pv}
  S.~J.~Sin, S.~Nakamura and S.~P.~Kim,
  ``Elliptic Flow, Kasner Universe and Holographic Dual of RHIC Fireball,''
  JHEP {\bf 0612}, 075 (2006)
  [arXiv:hep-th/0610113].

\bibitem{Herzog:2006gh}
  C.~P.~Herzog, A.~Karch, P.~Kovtun, C.~Kozcaz and L.~G.~Yaffe,
  ``Energy loss of a heavy quark moving through N = 4 supersymmetric
  Yang-Mills plasma,'' JHEP {\bf 0607}, 013 (2006)
  [arXiv:hep-th/0605158]

\bibitem{Gubser-drag}
  S.~S.~Gubser, ``Drag force in AdS/CFT,''
  Phys.\ Rev.\  D {\bf 74}, 126005 (2006)
  [arXiv:hep-th/0605182].

\bibitem{Wu:2014gla}
  Y.~Wu, D.~Hou and H.~c.~Ren,
  ``Some Comments on the Holographic Heavy Quark Potential in a Thermal Bath,''
  arXiv:1401.3635 [hep-ph].


\bibitem{Li:2014dsa}
  D.~Li, S.~He and M.~Huang,
  ``Temperature dependent transport coefficients in a dynamical holographic QCD model,''
  JHEP {\bf 1506} (2015) 046
  [arXiv:1411.5332 [hep-ph]].

\bibitem{Li:2014hja}
  D.~Li, J.~Liao and M.~Huang,
  ``Enhancement of jet quenching around phase transition: result from the dynamical holographic model,''
  Phys.\ Rev.\ D {\bf 89}, no. 12, 126006 (2014)
  [arXiv:1401.2035 [hep-ph]].


\bibitem{Li:2011hp}
  D.~Li, S.~He, M.~Huang and Q.~S.~Yan,
  ``Thermodynamics of deformed AdS$_5$ model with a positive/negative quadratic correction in graviton-dilaton system,''
  JHEP {\bf 1109} (2011) 041
  [arXiv:1103.5389 [hep-th]].


\bibitem{Cai:2012xh}
  R.~G.~Cai, S.~He and D.~Li,
 ``A hQCD model and its phase diagram in Einstein-Maxwell-Dilaton system,''
  JHEP {\bf 1203} (2012) 033
  [arXiv:1201.0820 [hep-th]].




\bibitem{DeWolfe:2010he}
  O.~DeWolfe, S.~S.~Gubser and C.~Rosen,
  ``A holographic critical point,''
  Phys.\ Rev.\ D {\bf 83} (2011) 086005
  [arXiv:1012.1864 [hep-th]].


\bibitem{DeWolfe:2011ts}
  O.~DeWolfe, S.~S.~Gubser and C.~Rosen,
  ``Dynamic critical phenomena at a holographic critical point,''
  Phys.\ Rev.\ D {\bf 84} (2011) 126014
  doi:10.1103/PhysRevD.84.126014
  [arXiv:1108.2029 [hep-th]].

\bibitem{Knaute:2017opk}
  J.~Knaute, R.~Yaresko and B.~K\"{a}mpfer,
  ``Holographic QCD phase diagram with critical point from Einstein-Maxwell-dilaton dynamics,''
  Phys.\ Lett.\ B {\bf 778} (2018) 419
  doi:10.1016/j.physletb.2018.01.053
  [arXiv:1702.06731 [hep-ph]].

\bibitem{Critelli:2017euk}
  R.~Critelli, R.~Rougemont and J.~Noronha,
  ``Homogeneous isotropization and equilibration of a strongly coupled plasma with a critical point,''
  JHEP {\bf 1712} (2017) 029
  [arXiv:1709.03131 [hep-th]].

\bibitem{Li:2016smq}
  D.~Li and M.~Huang,
  ``Chiral phase transition of QCD with $N_f=2+1$ flavors from holography,''
  JHEP {\bf 1702} (2017) 042
  [arXiv:1610.09814 [hep-ph]].



\bibitem{Fang:2015ytf}
  Z.~Fang, S.~He and D.~Li,
  ``Chiral and Deconfining Phase Transitions from Holographic QCD Study,''
  Nucl.\ Phys.\ B {\bf 907} (2016) 187
  [arXiv:1512.04062 [hep-ph]].

\bibitem{Li:2017ple}
  Z.~Li, Y.~Chen, D.~Li and M.~Huang,
  ``Locating the QCD critical end point through the peaked baryon number susceptibilities along the freeze-out line,''
  Chin.\ Phys.\ C {\bf 42} (2018) no.1,  013103
  [arXiv:1706.02238 [hep-ph]].

\bibitem{He:2013qq}
  S.~He, S.~Y.~Wu, Y.~Yang and P.~H.~Yuan,
  ``Phase Structure in a Dynamical Soft-Wall Holographic QCD Model,''
  JHEP {\bf 1304} (2013) 093
  [arXiv:1301.0385 [hep-th]].

\bibitem{Yang:2014bqa}
  Y.~Yang and P.~H.~Yuan,
  ``A Refined Holographic QCD Model and QCD Phase Structure,''
  JHEP {\bf 1411} (2014) 149
  [arXiv:1406.1865 [hep-th]].

\bibitem{Li:2017tdz}
  M.~W.~Li, Y.~Yang and P.~H.~Yuan,
  ``Approaching Confinement Structure for Light Quarks in a Holographic Soft Wall QCD Model,''
  Phys.\ Rev.\ D {\bf 96} (2017) no.6,  066013
  [arXiv:1703.09184 [hep-th]].

\bibitem{Chen:2017cyc}
  Y.~Chen, M.~Huang and Q.~S.~Yan,
  ``Gravitation waves from QCD and electroweak phase transitions,''
  JHEP {\bf 1805} (2018) 178
  [arXiv:1712.03470 [hep-ph]].

\bibitem{Kajantie:2011nx}
  K.~Kajantie, M.~Krssak, M.~Vepsalainen and A.~Vuorinen,
  ``Frequency and wave number dependence of the shear correlator in strongly coupled hot Yang-Mills theory,''
  Phys.\ Rev.\ D {\bf 84} (2011) 086004
  [arXiv:1104.5352 [hep-ph]].

\bibitem{criticalchiral}
  J.~Chen, S.~He, M.~Huang and D.~Li,
  ``Critical exponents of finite temperature chiral phase transition in soft-wall AdS/QCD models,''
  arXiv:1810.07019 [hep-ph].


\end{thebibliography}
\end{document}